# Frequency ratio measurement of $^{171}$Yb and $^{87}$Sr optical lattice clocks

**Daisuke Akamatsu,* Masami Yasuda, Hajime Inaba, Kazumoto Hosaka, Takehiko Tanabe, Atsushi Onae, and Feng-Lei Hong**

*National Metrology Institute of Japan (NMIJ), National Institute of Advanced Industrial Science and Technology(AIST), Central 3, 1-1-1 Umezono, Tsukuba, Ibaraki 305-8563, Japan*
*d-akamatsu@aist.go.jp*

**The frequency ratio of the $^1S_0(F=1/2)$-$^3P_0(F=1/2)$ clock transition in $^{171}$Yb and the $^1S_0(F=9/2)$-$^3P_0(F=9/2)$ clock transition in $^{87}$Sr is measured by an optical-optical direct frequency link between two optical lattice clocks. We determined the ratio ($\nu_{Yb}/\nu_{Sr}$) to be 1.207 507 039 343 340 4(18) with a fractional uncertainty of $1.5 \times 10^{-15}$. The measurement uncertainty of the frequency ratio is smaller than that obtained from absolute frequency measurements using the International Atomic Time (TAI) link. The measured ratio agrees well with that derived from the absolute frequency measurement results obtained at NIST and JILA, Boulder, CO using their Cs-fountain clock. Our measurement enables the first international comparison of the frequency ratios of optical clocks, and we obtained a good agreement between the two measured ratios with an uncertainty smaller than the TAI link. The measured frequency ratio will be reported to the International Committee for Weights and Measures for a discussion related to the redefinition of the second.**

## 1. Introduction

Recent developments in ultra precise laser technology have provided the possibility for time keeping at the $10^{-18}$ level [1-3]. Single ion clocks have exhibited a systematic uncertainty of $8.6 \times 10^{-18}$ [1], and enabled a tabletop demonstration of a relativistic gravitational effect [4]. While the ion clock uses a single quantum absorber for an atomic reference, an optical lattice clock uses thousands of atoms [5]. Thanks to the small quantum projection noise of an optical lattice clock, its stability can surpass that of an ion clock. Very recently a stability of $1.6 \times 10^{-18}$ has been achieved with an Yb optical lattice clock [2] and an accuracy of $6.4 \times 10^{-18}$ has been demonstrated with a Sr optical lattice clock [3]. Together with ion clocks, optical lattice clocks are now promising candidates of a new definition of the second.

In relation to redefining the second, absolute frequency measurements of optical lattice clocks have been reported from several institutes at the $10^{-15}$-$10^{-16}$ uncertainty level [6-14]. The measurement uncertainty is limited by the current frequency standards that realize the definition of the second. In institutes where Cs fountain clocks are available, the uncertainty of the absolute frequency measurements is usually at the $10^{-16}$ level. However, in most institutes, the absolute frequency measurements are based on the International Atomic Time (TAI) link and usually have an uncertainty of $10^{-15}$. Since the reproducibility of optical clocks can surpass that of Cs atomic clocks, the frequency ratio measurement of optical clocks by using an optical-optical frequency comparison can surpass the accuracy of the definition of the second. Consequently, the uncertainty of the frequency ratio measurement is limited by the uncertainty of the optical clocks. The results of an optical frequency ratio measurement would also enable us to cross check the absolute frequency measurement results by comparing the ratios measured in different institutes. Furthermore, after the redefinition of the second, the information on the frequency ratio could be used to realize the new SI second by using other optical clocks. The frequency ratio measurements are, of course, also important for tests in fundamental physics, such as an examination of the temporal variation of the fine-structure constant [15, 16]. However, optical frequency ratio measurements have thus far only been conducted at two institutes with limited atomic species ($Al^+/Hg^+$ [16] and $Ca^+/Sr$ [17]).

In this paper, we report the demonstration of a direct optical frequency comparison of $^{87}Sr$ and $^{171}Yb$ optical lattice clocks. We use only one narrow linewidth laser emitting at 1064 nm (master laser), whose frequency is stabilized to an ultra low expansion (ULE) optical reference cavity. We transfer the linewidth of the master laser to the clock lasers for Sr and Yb by using frequency combs with a broad servo bandwidth (the linewidth transfer method) [18]. The instability of the frequency ratio measurement is reduced by a factor of the square root two compared with when using separate ULE cavities for each clock laser. The frequency ratio of the $^1S_0(F=1/2)$-$^3P_0(F=1/2)$ clock transition in $^{171}Yb$ and the $^1S_0(F=9/2)$-$^3P_0(F=9/2)$ clock transition in $^{87}Sr$ ($\nu_{Yb}/\nu_{Sr}$) is determined to be 1.207 507 039 343 340 4(18) with a fractional uncertainty of $1.5\times10^{-15}$. The uncertainty of the frequency ratio measurement is more than twice smaller than that of the absolute frequency measurements using the TAI link in our institutes [11,12]. The obtained frequency ratio of Yb and Sr agrees with the calculated value using the previously reported absolute frequency measurements [6,13]. The ratio measurement provided a cross check of our absolute frequency measurements and increased the reliability of our optical lattice clocks. Our measurement scheme using laser linewidth transfer would allow clock laser noise cancellation with a synchronous measurement and enable us to realize frequency ratio measurements beyond the Dick limit [19].

## 2. Experimental setup

The experimental setup is shown schematically in Fig.1. An Nd:YAG laser operating at 1064 nm (master laser) was used to prepare the clock lasers for an $^{87}Sr$ optical lattice clock (Sr OLC) and a $^{171}Yb$ optical lattice clock (Yb OLC) by the linewidth transfer method. The master laser was stabilized to a high-finesse ULE cavity by the Pound-Drever-Hall technique. The instantaneous linewidth of the master laser was less than 3.5 Hz. The linewidth including frequency jitter at an averaging time of several seconds was 10 Hz.

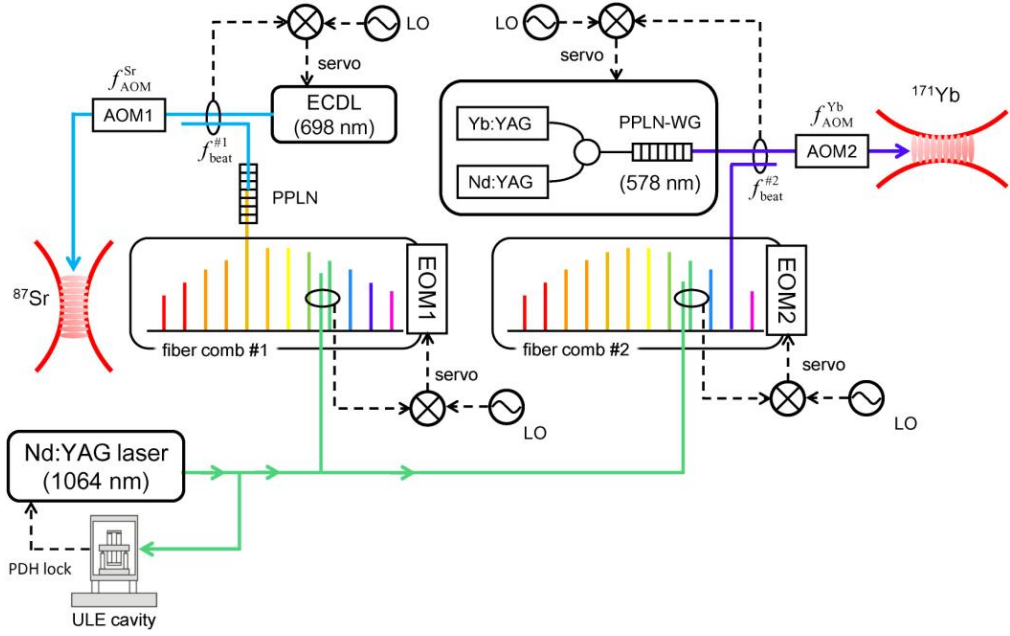

Fig. 1: Schematic diagram of experimental setup. The two optical lattice clocks (OLCs) have different lattice geometries: Sr OLC uses a vertical optical lattice and Yb OLC uses a horizontal optical lattice. ECLD, extended cavity laser diode; PPLN-WG, periodically poled lithium niobate waveguide; ULE cavity, ultra-low-expansion cavity; PDH lock, Pound-Drever-Hall lock; EOM, electro-optic modulator; AOM, acousto-optic modulator. Local oscillators (LOs) are microwave sources.

In this experiment, we employed fiber-type optical frequency combs (fiber combs) with a broad servo bandwidth for linewidth transfer (comb #1 and comb #2 in Fig.1). The oscillator cavity of the combs includes an electro-optic modulator (EOM) for a broad servo bandwidth (approximately 1.3 MHz). The fiber combs are described in detail in [18, 20, 21]. A heterodyne beat was detected between the master laser and the frequency component of comb #1(#2) $f_{beat}^{\#1(\#2)}$ and then stabilized so that $f_{beat}^{\#1}(-f_{beat}^{\#2}) = 30$ MHz using the intracavity EOM1(EOM2). The carrier-envelope offset frequency was detected using a common-path $f$-$2f$ interferometer and stabilized so that $f_{ceo}^{\#1} = -f_{ceo}^{\#2} = 30$ MHz. As a result, the repetition rates of the combs were stabilized so that $f_{rep}^{\#1} = f_{rep}^{\#2} = 43.4$ MHz. The linewidth of the comb components was reduced to Hz level (narrow-linewidth fiber combs). The observed beat linewidth between these two combs was 30.1 mHz, which was the measurement limit of the spectrum analyzer, and the frequency stability of the beat was $3\times10^{-16}$ at a 1-s averaging time [21].

The clock laser for the Sr (Yb) OLC was prepared by transferring the linewidth of the master laser to the lasers at 698 nm (578 nm) through the narrow-linewidth fiber combs. A 698 nm light was generated by an extended cavity diode laser. Second-harmonic comb components around 698 nm were generated by passing the fundamental comb through a periodically poled lithium niobate (PPLN) crystal. The laser was phase-locked to the fiber comb by detecting and stabilizing a heterodyne beat note between the second-harmonic comb component and the clock laser. The tight phase-locking of the fiber comb to the master laser and the clock laser to the fiber comb allowed the linewidth of the master laser to be transferred to the clock laser. We note that comb #1 was also used to narrow the linewidth of a 689 nm laser for the $2^{nd}$ stage laser cooling of the Sr OLC [22]. A 578 nm light source was generated by the sum-frequency generation of a Nd:YAG laser at 1319 nm and an Yb:YAG laser at 1030 nm using a PPLN waveguide (PPLN-WG)[23]. The clock laser for the Yb OLC was stabilized to fiber comb #2 in a similar manner.

The Sr OLC at the National Metrology Institute of Japan (NMIJ) is operated with spin-polarized $^{87}$Sr atoms. $^{87}$Sr atoms were slowed and magneto-optically trapped on a $^1S_0$-$^1P_1$ transition at 461 nm [24]. Then the atoms were further cooled to a few μK by using the inter-combination transition $^1S_0$-$^3P_1$ [22]. The cooled atoms were loaded into a vertically one-dimensional optical lattice operating at $f_{magic}$ = 368.5545 THz. The lattice laser frequency was monitored during the measurement with a calibrated wavelength meter and its fluctuation was kept within 300 MHz. The trap depth of the lattice laser was 28 $E_r$ (the lattice photon recoil energy $Er/k_B$=165 nK). The atoms were optically pumped to the Zeeman sublevel $m_F = \pm 9/2$ by using a weak light resonant with $^1S_0$-$^3P_1$. The typical spectrum of a Zeeman component is shown in Fig. 2(a) for an interrogation time of 40 ms. A Fourier limited linewidth of 22 Hz was observed with a large excitation probability resulting a good signal-to-noise ratio for the spectrum.

The Yb OLC at the NMIJ is operated with non spin-polarized $^{171}$Yb atoms. The Yb OLC has been described previously [12]. After 1$^{st}$ stage cooling by a laser at 399 nm, the atoms were further cooled to a few tens of μK by using the inter-combination transition $^1S_0$-$^3P_1$ [25]. The cooled atoms were loaded into a horizontally oriented one-dimensional optical lattice operating at $f_{magic}$ = 394 798. 337 GHz. The lattice laser frequency was stabilized to a third optical frequency comb (not shown in Fig.1) using a delay line lock technique with an uncertainty of 10 MHz. The trap depth of the lattice laser was 500 $E_r$ (the lattice photon recoil energy $Er/k_B$=97 nK). The typical spectrum of the Yb clock transition is shown in Fig. 2(b) for a 40 ms interrogation pulse. The observed linewidth was broader than the Fourier limited linewidth because of the saturation broadening. Both OLCs were located in the same laboratory room.

For the clock operation, the clock laser was frequency-stabilized to both stretched states ($m_F = \pm 9/2$ for Sr OLC; $m_F = \pm 1/2$ for Yb OLC) using an acousto-optic modulator (AOM1 for Sr OLC; AOM2 for Yb OLC) to eliminate the first-order Zeeman shift. The clock laser alternately probed the high- and low-frequency points at the full width of half maximum of the stretched state spectrum. A single probe cycle was 2 s for the Sr OLC and 1 s for the Yb OLC. The RF frequency driving the AOMs was controlled to equalize the excitation probabilities such that $p_+^{FWHM}(\pm m_F) = p_-^{FWHM}(\pm m_F)$. The average RF frequency ($f_{AOM}^{Sr(Yb)}$) determined the center frequency of the clock transition without the first-order Zeeman shift. We simultaneously operated the Sr and Yb OLCs for the frequency ratio measurement.

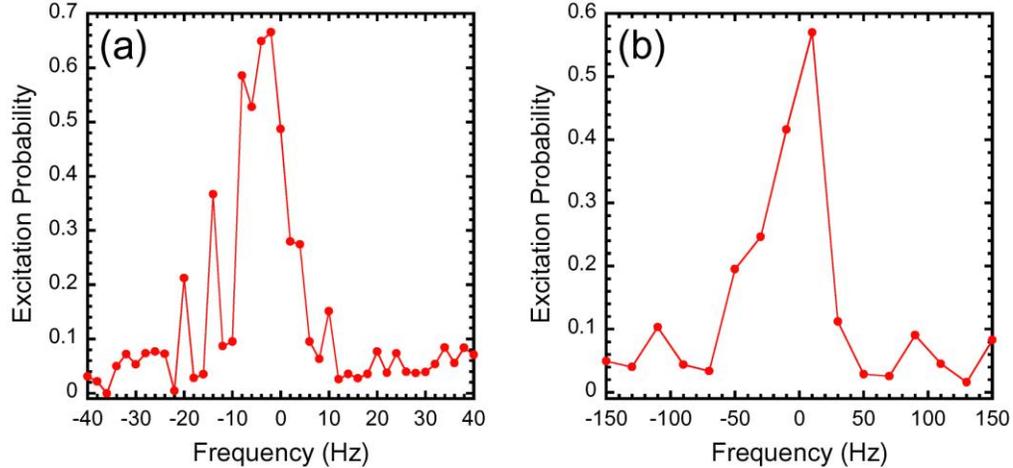

Fig. 2: Typical stretched state spectrum of the clock transition of (a) the Sr OLC and (b) the Yb OLC. The interrogation pulse durations were 40 ms in both cases.

3. **Experimental results**

The absolute frequency of the Sr (Yb) can be calculated by
$$\nu_{Sr} = n_{Sr} f_{rep}^{\#1} + f_{CEO}^{\#1} + f_{beat}^{\#1} + f_{AOM}^{Sr}$$
$$\nu_{Yb} = n_{Yb} f_{rep}^{\#2} + f_{CEO}^{\#2} + f_{beat}^{\#2} + f_{AOM}^{Yb}.$$
As we described in the experimental setup section,
$$f_{rep}^{\#1} = f_{rep}^{\#2} = f_{rep}$$
$$f_{CEO}^{\#1} = -f_{CEO}^{\#2} = f_{CEO}.$$
Therefore, the frequency ratio measurement of the clock transitions can be derived as
$$\frac{\nu_{Yb}}{\nu_{Sr}} = r - \frac{(1+r)f_{CEO} - f_{beat}^{\#2} - f_{AOM}^{Yb} + r(f_{beat}^{\#1} + f_{AOM}^{Sr})}{\nu_{Sr}},$$
where $r = n_{Yb}/n_{Sr} \approx 1.2$. Since the radio frequencies are divided by the optical frequency in the second term, the term is of the order of $10^{-6}$. Therefore the measurement uncertainty of the frequency ratio would be at the $10^{16}$ level even with the $10^{-10}$ measurement uncertainties for the radio frequencies [17].

Figure 3(a) shows 4 frequency ratio measurement results, where the corrections are included. Figure 3(b) shows a typical Allan standard deviation of the measured frequency ratio (# 3 in Fig. 3(a)). The Allan standard deviation was $9.9\times10^{-15}$ for an 8 s averaging time. The short-term stability is determined by that of the master laser. We note that if we use the clock lasers own ULE cavities, the short-term stability would become worse by a factor of the square root of two assuming the cavities have the same specifications as the cavity for the master laser. The Allan standard deviation is improved to $2.0\times10^{-15}$ after an 840 s averaging time. Since the experiment was a direct comparison of the optical frequencies, the obtained Allan deviation was much smaller than that of the microwave reference used in the absolute frequency measurement. The error bars were determined by the Allan standard deviation at the longest averaging time for each measurement (e.g. the uncertainty for measurement # 3 in Fig. 3(a) was calculated by using the Allan standard deviation in Fig. 3(b)).

The corrections and uncertainties for the frequency ratio measurement were calculated from those for the Sr and Yb OLCs (Table 1). The largest corrections for both the Sr and Yb OLCs were due to the blackbody radiation (BBR) shifts. The BBR corrections for the Sr OLC (Yb OLC) was estimated to be $55.3(2.3)\times10^{-16}$ ($25.2(2.0)\times10^{-16}$) for the measured temperature of the vacuum chamber T = 303(3) K (T=302(3)K). The evaluated ac Stark shift including the hyperpolarizability for the Sr OLC (Yb OLC) by the lattice laser was $0.0(3.6)\times10^{-16}$ ($2.1(2.4)\times10^{-16}$). The light shifts induced by the clock laser for the Sr OLC and Yb OLC were estimated to be $-0.2(2)\times10^{-16}$ and $0.8(2)\times10^{-16}$, respectively, using the measured clock laser intensities (0.3 mW/cm$^2$ for Sr; 2.3 mW/cm$^2$ for Yb). The second-order Zeeman shifts caused by bias magnetic fields (91 µT for Sr; 170 µT for Yb) were estimated to be $-4.3(1.1)\times10^{-16}$ and $-3.8(4)\times10^{-16}$, respectively. The collision shifts for the Sr OLC was estimated to be $0.0(2.1)\times10^{-16}$ by an interleaved method. The collisional shift for the Yb OLC was estimated to be $-0.3(2.6)\times10^{-16}$ using data obtained at NIST and our number of atoms, assuming a similar trap volume. We neglected the gravitational shift between the two clocks due to the height difference of 56(5) mm.

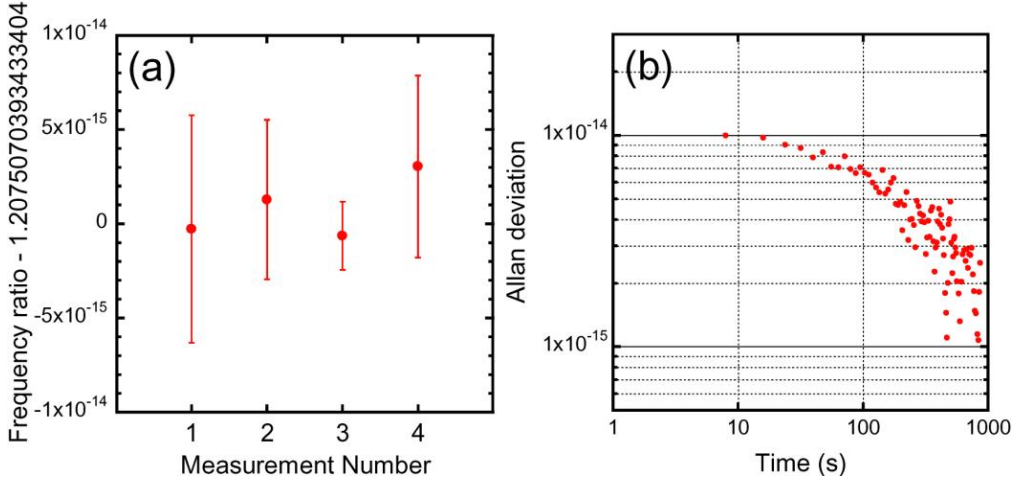

Fig.3: (a) Frequency ratio measurements of Sr and Yb optical lattice clocks. The error bars are statistical. Data shown in this figure include the systematic corrections. (b) Allan standard deviation of the frequency ratio measurement for the measurement number 3 in Fig.3 (a).

In the experiment, we found a relatively large servo error for the Sr OLC. In the clock laser frequency stabilization to the clock transition, the excitation probabilities for high- and low-frequency points $p_+^{FWHM}(m_F = \pm 9/2)$, $p_-^{FWHM}(m_F = \pm 9/2)$ were not the same value because of the clock laser linear drift and the weak feedback servo gain, which results in a residual frequency offset for the Sr OLC. We determined the frequency offset for each measurement from the difference between the averaged excitation probabilities using the lineshape shown in Fig. 2(a). The uncertainty was determined from the statistical deviation of the mean of the excitation probability difference.

The frequency ratio of Yb/Sr was determined from the weighted mean of the four measurements shown in Fig. 4(a) to be 1.207 507 039 343 340 4(18). The fractional uncertainty of the frequency ratio $1.5\times10^{-15}$ was determined by combining the systematic uncertainty ($8.1\times10^{-16}$) with the statistical uncertainty of the measurements ($1.3\times10^{-15}$).

Table 1: The fractional uncertainties and corrections of the Sr OLC, Yb OLC, and the frequency ratio measurement. The asterisks represent the value depending on the measurements.

| Effect | Correction ($\times10^{-16}$) | | | Uncertainty ($\times10^{-16}$) | | |
|---|---|---|---|---|---|---|
| | Sr | Yb | Yb/Sr | Sr | Yb | Yb/Sr |
| Blackbody radiation | 55.3 | 25.2 | -36.3 | 2.3 | 2.0 | 3.6 |
| AC Stark (lattice) | 0.0 | -2.1 | -2.5 | 3.6 | 2.4 | 5.2 |
| AC Stark (probe) | 0.2 | -0.8 | -1.2 | 0.2 | 0.2 | 0.3 |
| 2nd order Zeeman | 4.3 | 3.8 | -0.6 | 1.1 | 0.4 | 1.3 |
| Collision | 0.0 | 0.3 | 0.3 | 2.1 | 2.6 | 4.0 |
| Servo error | * | 0.0 | * | 2.4 | 0.0 | 2.8 |
| ***Sr/Yb systematics total*** | | | * | | | 8.1 |

4. **Discussion and conclusion**

The precise frequency ratio of optical clocks can also be derived from absolute frequency measurements where a Cs fountain clock is used. The absolute frequencies of a Sr OLC and an Yb OLC have been measured and found to be 429 228 004 229 873.65(37) Hz (fractional uncertainty $8.6\times10^{-16}$) and 518 295 836 590 865.2(7) Hz (fractional uncertainty $1.4\times10^{-15}$),

respectively, by using one of the best Cs fountain clocks (NIST F-1) in Boulder, CO [6, 13]. From these results, the frequency ratio was calculated to be 1.207 507 039 343 339 7(20) with a fractional uncertainty of $1.6\times10^{-15}$. Our experimental result 1.207 507 039 343 340 4(18) agrees with the Boulder result within their uncertainties. We would like to emphasize that the total measurement time was only 12000 s for this experiment, which is much shorter than the absolute frequency measurements ($>10^5$ s). This is the first international comparison of the frequency ratios of optical clocks. The agreement between the Tsukuba and Boulder values is $6\times10^{-16}$, which is much smaller than the TAI link ($\sim3\times10^{-15}$). Since the TAI link is used for absolute frequency measurements of optical clocks at various metrology institutes, the comparison of the frequency ratio provides a good cross check of absolute frequency measurement results. Furthermore, the frequency ratio comparison can also be used to verify the uncertainty evaluation results for each clock.

The uncertainty of the measured frequency ratio is limited by the statistical uncertainty in the experiment. With a longer averaging time, the statistical uncertainty could be reduced and the measurement uncertainty would eventually be limited by the systematic uncertainties of the OLCs. Uncertainty evaluations at the $10^{-17}$-$10^{-18}$ level for Sr optical lattice clocks have been demonstrated recently at several laboratories [3, 7, 9]. Therefore, the uncertainty of the frequency ratio measurement is expected to reach the $10^{-17}$-$10^{-18}$ level. For such an experiment, synchronized interrogation [19] would be a powerful tool for improving the short-term stability and hence reduce the total measurement time.

Optical clocks are being evaluated by the International Committee for Weights and Measures (CIPM) with a view to redefining the second. Together with the absolute frequencies, the frequency ratios measured in different laboratories will be reported to the CIPM for discussions related to the redefinition of the second. As with the absolute frequencies, CIPM may also recommend frequency ratios of optical lattice clocks based on the reported results. A matrix table of frequency ratios of various optical clocks will be established for cross checking measurement results. This may reveal any undiscovered errors with the optical clocks. After the redefinition of the second, the matrix table should be used to realize the second with optical clocks other than the new definition. Our measured frequency ratio will be reported to the CIPM via the next Consultative Committee for Time and Frequency scheduled for 2015, and will contribute to discussions regarding the recommendation of the frequency ratio of optical clocks, which is important for the redefinition of the second.

**Acknowledgments**
This research receives support from the JSPS through its FIRST Program and JSPS KAKENHI Grant Numbers 13222778 and 23540472.